\documentclass{nature}
\usepackage{graphicx}
\makeatletter
\let\saved@includegraphics\includegraphics
\AtBeginDocument{\let\includegraphics\saved@includegraphics}
\renewenvironment*{figure}{\@float{figure}}{\end@float}
\makeatother
\usepackage{dcolumn}
\usepackage{color}
\usepackage{soul}%
\usepackage{ulem}%
\usepackage{bm}%
\usepackage{braket}
\usepackage{amsmath}
\usepackage{float}
\usepackage{xcolor}
\usepackage{enumerate}
\usepackage[shortlabels]{enumitem}

\begin{document}

\title{Robust and efficient algorithms for high-dimensional black-box quantum optimization}

\author{Zhaoqi Leng$^1$, Pranav Mundada$^2$,  Saeed Ghadimi $^3$ \& Andrew Houck $^2$}

\maketitle
\begin{affiliations}
 \item Department of Physics, Princeton University.
 \item Department of Electrical Engineering, Princeton University.
 \item Department of Operations Research and Financial Engineering, Princeton University.
\end{affiliations}

\begin{abstract}
Hybrid quantum-classical optimization using near-term quantum technology is an emerging direction for exploring quantum advantage in high-dimensional systems. However, precise characterization of all experimental parameters is often impractical and challenging. A viable approach is to use algorithms that rely only on black-box inference rather than analytical gradients. Here, we combine randomized perturbation gradient estimation with adaptive momentum gradient updates to create the AdamSPSA and AdamRSGF algorithms. We prove the asymptotic convergence of our algorithms in a convex setting, and we benchmark them against other gradient-based optimization algorithms on non-convex optimal control tasks. Our results show that these new algorithms accelerate the convergence rate, decrease the variance of loss trajectories, and efficiently tune up high-fidelity (above 99.9\%) Hann-window single qubit gates from trivial initial conditions with twenty variables.
\end{abstract}
\maketitle
\begin{figure}[!htbbp]
\includegraphics[width=\linewidth, trim= 2cm 0.5cm 2cm 0.5cm,clip]{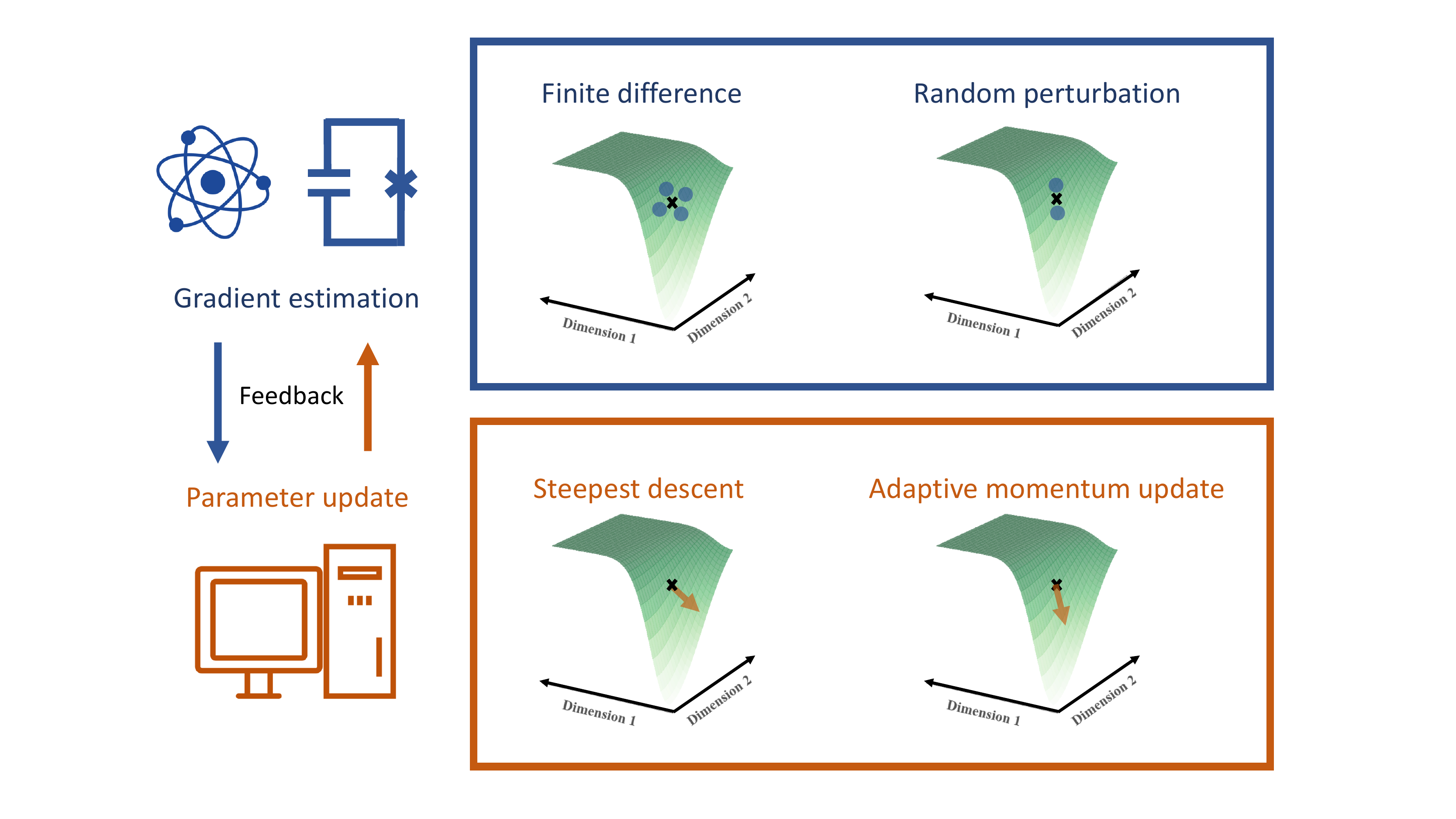}
\caption{\textbf{Schematic for gradient-based black-box optimization}. Hybrid quantum-classical optimization contains two stages: gradient estimation, in which the function evaluation runs on a quantum computer (either with real atoms or artificial atoms), and parameter update, which runs on a classical computer. Green surfaces are synthesis loss landscapes with a two-dimensional domain. The black cross marks are the locations of interest. Blue dots represent function evaluations for estimating the gradient. The cost of gradient estimation for finite difference is linearly proportional to the dimension of the domain, whereas the cost of random perturbation is independent of the dimension of the domain. Orange arrows represent the directions of parameter update. Steepest descent moves along the direction of the estimated gradient while adaptive momentum update depends on the history of the previous estimated gradients and regulates the step size based on the history of squared estimated gradients. The two stages are repeated to lower a loss function until a termination condition is reached. }
\label{fig:fig0}
\end{figure}

Closed-loop feedback between quantum and classical systems has applications in the fields of optimal quantum control, quantum chemistry, and quantum machine learning, because of potential benefits from quantum advantage\cite{kelly2014optimal,rol2017restless,kandala2017hardware, zhu2018training, king2018observation, o2016scalable, otterbach2017unsupervised, hu2019quantum, kokail2018self, Cong2019}.   Gradient-based algorithms are a common approach to optimization: they are efficient for high-dimensional problems and theoretically guaranteed to find the global optimum of convex objective functions. Typically, gradients can be extracted from an analytical model, but constructing such a model requires precise understanding of the system of interest. As noisy intermediate-scale quantum (NISQ) systems expand in quantum volume and progress toward quantum supremacy \cite{preskill2012quantum, debnath2016demonstration, boixo2018characterizing, preskill2018quantum, cross2018validating, king2018observation,neill2018blueprint, wright2019benchmarking}, the experimental uncertainties (e.g. unwanted signal distortion, interactions, on-chip microwave cross talk, and effects from higher energy levels of the qubits) become increasingly challenging to analytically capture.\\
\indent When an analytical gradient cannot be derived, gradient-based closed-loop feedback requires gradient estimation and parameter updates, as shown in Fig \ref{fig:fig0}. These two steps are repeated until an objective function $f(\bm{\theta})$ converges or another termination condition is reached. Unfortunately, using simple finite-difference methods to perform gradient estimation is computationally expensive. In addition, the parameter update rule for simple gradient descent is vulnerable to the effects of stochastic noise and shallow local minima. These limits on typical gradient-based methods pose challenges to closed-loop optimization on quantum systems. Therefore, it is crucial to look beyond typical gradient-based methods to develop algorithms that are robust to noisy gradients and shallow local minima and are efficient for high-dimensional optimization. \\
\indent Specifically, when optimizing an objective function $f(\bm{\theta})$ via a black-box interface with a quantum system, only zeroth-order information $f(\bm{\theta})$ is available through system measurements; thus, the gradient $\nabla_{\bm{\theta} }f(\bm{\theta})$ cannot be analytically calculated without precise knowledge of the system of interest. To perform gradient-based parameter updates, the gradients need to be estimated with zeroth-order information as in the finite-difference method. When the estimated gradient is noisy and biased, a robust gradient descent algorithm is critical for updating parameter $\bm{\theta}$. Optimization routines---such as heavy-ball momentum and adaptive learning rate---have been developed by the deep learning community to increase convergence rates and lower loss function values \cite{polyak1964some, rumelhart1988learning, sutskever2013importance, wilson2016lyapunov, duchi2011adaptive, zeiler2012adadelta, tieleman2012lecture, dauphin2015equilibrated, dozat2016incorporating, reddi2019convergence}. However, these methods rely on first-order information, which can be efficiently computed with the backpropagation algorithm in deep learning\cite{rumelhart1988learning}. Studies on improving optimization using only zeroth-order information, specifically for quantum computing purposes, are still lacking but important. \\
\indent In this paper, we propose new black-box quantum optimization algorithms that combine momentum and adaptive learning rate scheduling with random-perturbation-based gradient estimation. We experimentally benchmark these algorithms against other gradient-based algorithms on quantum optimal control of a single transmon qubit. Compared to standard finite difference and random-perturbation-based algorithms, our algorithms show significant improvement for twenty-dimensional optimization, robustness against local minima, and resilience to experimental noise. We further demonstrate that the gradient-based black-box optimization algorithms can be used to tune up single qubit gates with high fidelity given noisy and biased gradient estimation.
\section*{Results}
\subsection{Stochastic gradient descent.} Stochastic gradient-based optimization can be framed as a generalized Robbin-Monro algorithm where optimization parameters $\theta$ are updated iteratively by their gradients \cite{robbins1951stochastic}. The update rule can be written as the following:
\begin{equation}
\begin{split}
\bm{\theta}_{t+1} &= \bm{\theta}_{t}  - a_t \hat{\bm{g}}(\bm{{\theta_{t}}})\\
&=\bm{\theta_{t}} - a_t\big( \bm{g}(\bm{\theta_{t}}) +  \bm{b}_t(\bm{\theta_{t}}) +  \bm{\bm{e}}_t(\bm{\theta_{t}})\big),
\end{split}
\end{equation}
where $a_t$ is the learning rate at training step t. The gradient estimator $\hat{\bm{g}}(\bm{\theta_{t}})$ differs from the true gradient $\bm{g}(\bm{\theta_{t}})$ by two additional components: the bias of the gradient $ \bm{b}_t(\bm{\theta_{t}}) = E[\hat{\bm{g}}(\bm{\theta_{t}}) - \bm{g}(\bm{\theta_{t}})]$  and the stochastic noise $\bm{e}_t(\bm{\theta_{t}}) = \hat{\bm{g}}(\bm{\theta_{t}}) - E[\hat{\bm{g}}(\bm{\theta_{t}})]$. In the rest of this section, we will first describe general gradient estimation involving finite-difference methods. Then, we will describe more efficient versions of these general schemes---randomized perturbation gradient estimation---which will be used in our new algorithms (AdamSPSA and AdamRSGF). Finally, we will introduce the update rules for AdamSPSA and AdamRSGF, which incorporate the estimated momentum and the estimated gradient information into the learning rate $a_t$.
\subsection{General gradient estimation.} In black-box optimization, the gradient estimator $\hat{\bm{g}}(\bm{\theta}_{t})$ can be computed with either Kiefer-Wolfowitz (KW) finite-difference algorithms or smooth function  approx-imation\cite{kiefer1952stochastic, rockafellar1998rjb, yousefian2012stochastic, duchi2012randomized}. For KW-based algorithms, the $i^{th}$ component of the symmetric finite difference gradient is the following:
\begin{equation}
\begin{split} \label{eq:KW}
\hat{g}(\bm{\theta}_{t})^i = \frac{\hat{f}^+(\bm{\theta}_{t} + c_t \bm{e}^i) - \hat{f}^-(\bm{\theta}_{t} - c_t \bm{e}^i)}{2c_t},\\
\end{split}
\end{equation}
where $c_t$ is the size of a small perturbation, $e^i$ is a one-hot vector with 1 at the $i^{th}$ component, and $\hat{f}^+ = f^+ + \epsilon_t^+$ and  $\hat{f}^- = f^- + \epsilon_t^-$ are forward and backward function evaluations with stochastic noise $\epsilon^\pm$. Here, for a $p$-dimensional $\bm{\theta}_{t}$, the central difference gradient estimator  has stochastic noise $e_t(\bm{\theta}_{t}) = O(1/c_t)$, and it requires $2p$ function evaluations (on both sides of $\bm{\theta_t}$) giving a bias $b_t(\bm{\theta}_{t}) = O(c_t^2$)  \cite{fornberg1988generation}.\\
\indent In smooth function approximation, a function $f$ is convolved with a non-negative, measurable, bounded function $\psi(\bm{u}) $ that satisfies $\int \psi(\bm{u}) d\bm{u} = 1$ \cite{rockafellar1998rjb, yousefian2012stochastic, duchi2012randomized}. In Gaussian smoothing, $\bm{u}$ is drawn from a Gaussian distribution. The function approximation $\tilde{f}_{\bm{u}}$ and gradient $\hat{\bm{g}}$ are given by:
\begin{equation}
\begin{split}\label{eq:smooth}
\tilde{f}_{\bm{u}}(\bm{\theta}_t) &= \bigg(\frac{1}{2\pi}\bigg)^{p/2} \int \hat{f}(\bm{\theta}_t + c_t \bm{u})e^{-\frac{1}{2}||\bm{u}||^2} d\bm{u},\\
\hat{\bm{g}}(\bm{\theta}_{t}) &= \bigg(\frac{1}{2\pi}\bigg)^{p/2} \int \frac{\tilde{f}^+(\bm{\theta}_t + c_t \bm{u}) - \tilde{f}^-(\bm{\theta}_t)}{c_t} \bm{u} e^{-\frac{1}{2}||\bm{u}||^2} d\bm{u},
\end{split}
\end{equation}
where $\bm{u}$ is a $p$-dimensional standard Gaussian random vector and $c_t$ is the smoothing parameter \cite{nesterov2017random}. Contrary to KW-based algorithms, smooth function approximation involves single-sided gradient evaluation, which does not introduce additional bias; however, approximating the function via Gaussian convolution introduces $O(c_t^2)$ bias \cite{ghadimi2013stochastic}. The stochastic noise of the gradient using Gaussian smoothing is $e_t(\theta_{t}) = O(1/c_t)$. Thus, Gaussian smoothing results in the same orders of bias and stochastic noise as KW-based algorithms, even though it only has a single-sided function evaluation.
\subsection{Randomized perturbation gradient estimation.}
In a high-dimensional domain, the computational cost of the general gradient estimation methods shown above scales with the dimension of the optimization parameters $\bm{\theta}$. Our new algorithms will instead use randomized perturbation gradient estimation methods, which can alleviate linear growth of computational cost while keeping the bias $b_t =O( c_t^2)$ and variance $e_t$ = $O(1/c_t)$ of the gradient estimator the same orders of magnitude \cite{spall1992multivariate, ghadimi2013stochastic}. In particular, we will use simultaneous perturbation stochastic approximation (SPSA) and randomized stochastic gradient free (RSGF) methods, in which the gradient is estimated by perturbing the objective function using a random vector ($\bm{\Delta}$ or $\bm{u}$). A common choice of perturbation $\bm{\Delta}$ for SPSA is a Rademacher distribution where each value in $\bm{\Delta}$ has 0.5 probability of being +1 or -1. The gradient estimator for SPSA is given by:
\begin{equation}
\begin{split} \label{eq:SPSA}
\hat{g}(\bm{\theta}_{t})^i = \frac{\hat{f}^+(\bm{\theta}_{t} + c_t \bm{\Delta}_t) - \hat{f}^-(\bm{\theta}_{t} - c_t \bm{\Delta}_t)}{2c_t} \frac{1}{\bm{\Delta}_t^i}.\\
\end{split}
\end{equation}
\indent The gradient estimator for RSGF with Gaussian smoothing can be written as
\begin{equation}
\begin{split} \label{eq:RSGF}
\hat{g}(\bm{\theta}_{t})^i = \frac{\hat{f}^+(\bm{\theta}_{t} + c_t \bm{u}_t) - \hat{f}(\bm{\theta}_{t})}{c_t}\bm{u}_t^i,
\end{split}
\end{equation}
where $\bm{u}_t$ is a Gaussian random vector with mean 0 and identity covariance matrix. Both methods reduce the computational cost for the gradient by a factor of $p$. The two methods have similar forms because they both estimate the gradient by randomly perturbing all directions simultaneously, but they originate from different concepts: SPSA is based on KW finite-difference, whereas RSGF is based on smooth function approximation.\\
\indent  The noise and variance of the estimated gradients in equations \ref{eq:SPSA} and \ref{eq:RSGF} can be reduced by taking the mean of $N$ samplings: $\hat{\bm{g}}(\bm{\theta}_{t})^i = 1/N \sum_j \hat{\bm{g}}(\bm{\theta}_{t})^i_j$. In the experiments presented in the following sections, SPSA-based gradient estimation involved one sample with two function evaluations per gradient computation ($N=1$), and RSGF-based gradient estimation involved averaging two samples with one function evaluation per gradient computation ($N=2$). Thus, each algorithm takes the same number of function evaluations in total.
\subsection{Parameter update.}
Given an efficient yet noisy and biased gradient estimation method from Eqn \ref{eq:SPSA} or \ref{eq:RSGF}, a well-designed gradient descent rule is critical for robust parameter updates. We propose two new algorithms---AdamSPSA and AdamRSGF---which combine randomized perturbation gradient estimation (Eqn \ref{eq:SPSA}, \ref{eq:RSGF}) with adaptive momentum estimation, a technique popularized in the "Adam" algorithm\cite{kingma2014adam}. Adam is a first-order gradient descent algorithm in deep learning which directly uses first-order information, making it suitable for noisy and sparse gradients \cite{kingma2014adam}.  The update rule can be written as the following: 
\begin{equation}
\begin{split}
\hat{\bm{\theta}}_{t+1}^i &= \hat{\bm{\theta}}_{t}^i  - \frac{a_t}{\sqrt{\hat{\bm{v}}_t^i} + \delta} \hat{\bm{m}}_t^i,\\
\hat{\bm{m}}_t^i &= \frac{\beta_t \bm{m}_{t}^i}{\sum_{N=0}^t (1 - \beta_{t-1-N})\Pi_{i=0}^N \beta_{t-i}},\\
\hat{\bm{v}}_t^i &= \frac{\gamma_t \bm{v}_{t}^i}{\sum_{N=0}^t (1 - \gamma_{t-1-N})\Pi_{i=0}^N \gamma_{t-i} },\\
\bm{m}_t^i &= \beta_t \bm{m}_{t-1}^i + (1 - \beta_t)\hat{\bm{g}}_t^i,\\
\bm{v}_t^i &= \gamma_t \bm{v}^i_{t-1} + (1 - \gamma_t)(\hat{\bm{g}}_t^i)^2,\\
\end{split}
\end{equation}
\indent Contrary to the original Adam algorithm, the gradient $\hat{\bm{g}}_t$ of our new algorithms is evaluated using either SPSA or RSGF. Because SPSA and RSGF rely on zeroth-order information, they are noisy and biased. To combat this noise and bias, the update rule in Eqn 6 implements adaptive momentum estimation from Adam as follows. To update parameters $\bm{\theta}$, heavy-ball momentum $\hat{\bm{m}}_t$ is used instead of the gradient $\hat{\bm{g}}_t$, allowing the algorithm to escape local minima and averaging out the stochastic noise. In addition, the learning rate $a_t$ is adjusted based on the exponential moving average of squared estimated gradient $\hat{\bm{v}}_t$, thereby boosting the learning rate for the dimension with sparse gradient. To correct for the bias of exponential moving average, $\bm{m}_t$ and $\bm{v}_t$ are renormalized by weights $\beta_t$ and $\gamma_t$ respectively. Instead of using a constant learning rate $a_t$ and momentum coefficient $\beta_t$ (as in the original Adam algorithm\cite{kingma2014adam}), we require that $a_t$, $\beta_t$, and the perturbation size $c_t$  converge to zero to ensure asymptotic convergence in convex settings. It is important to note that the original Adam algorithm assumes that the unbiased gradient $\hat{\bm{g}}$ is available, whereas for black-box gradient estimation, the bias and noise of estimated gradients depend on the size of the perturbation. Therefore, the new algorithms need additional conditions to ensure convergence to an optimal value (proof in the supplement), as optimization may result in divergence, i.e. as $ c_t \to 0$, the bias $b_t$ converges to $0$ while the noise $e_t$ diverges.
\subsection{Experimental demonstration.}
Optimal control of a single transmon qubit is well-studied both theoretically and experimentally \cite{motzoi2009simple, chow2010optimized, lucero2010reduced, theis2016simultaneous}. Our focus here is to use single qubit optimal control as a test to evaluate the optimization algorithms. While previous experiments have shown single qubit derivative-removal-by-adiabatic-gate (DRAG) optimization with two variables in a locally convex setting \cite{kelly2014optimal, rol2017restless}, we benchmark our black-box optimization algorithms in a non-convex setting without prior knowledge of qubit leakage outside the computational space, signal distortion due to classical electronics, or a good initial condition for $(\bm{A}, \bm{B})$. Without loss of generality, we test the new algorithms on tuning a X$_{90}^{\text{Hann}}$ gate, which is parameterized as linear combinations of Hann windows in both the in-phase (I) and quadrature (Q) channels:
\begin{equation}
\begin{split}
I_{X_{90}} &= \sum_{i=1}^N A_i \big(1 - \cos\big(\frac{2\pi i t}{T}\big)\big), \\
Q_{X_{90}} &= \sum_{i=1}^N B_i \big(1 - \cos\big(\frac{2\pi i t}{T}\big)\big),
\end{split}
\end{equation}
where the duration of the pulse is $T=20$\,ns and each channel has 10 free parameters ($N = 10$). \\
\indent To benchmark the algorithms, we need to define an optimization loss function for tuning the X$_{90}^{\text{Hann}}$ gate. Simple loss functions using bootstrapping amplitude or Rabi amplitude as the only constraint will lead to arbitrary Bloch sphere axes. Such loss functions are valid for tuning a high fidelity $\tilde{X}_{90}^{\text{Hann}}$ gate and the corresponding orthogonal $\tilde{Y}_{90}^{\text{Hann}}$ gate. However, such a $\tilde{X}_{90}^{\text{Hann}}$ gate is hard to directly benchmark with the state of the art DRAG gate since its phase is arbitrary. Therefore, for the purpose of benchmarking, we enforce the orientation of the X$_{90}^{\text{Hann}}$ gate by providing longer 40\,ns X$_{90}^{\text{Ref}}$ and Y$_{90}^{\text{Ref}}$ reference gates (along I and Q channels) as additional constraints on the phase. These additional constraints increase the difficulty of tuning a X$_{90}^{\text{Hann}}$ gate yet are not necessary for the purpose of tuning high fidelity gates with arbitrary phases.\\
\indent We design an auxiliary optimization task to benchmark the algorithms, which is later incorporated into high-fidelity gate optimization. The qubit is first initialized in $\ket{\psi}^X = 1/\sqrt{2} \ket{0} -i/\sqrt{2} \ket{1}$ by applying a reference X$_{90}^{\text{Ref}}$ gate. Then, the X$_{90}^{\text{Hann}}$ gate is applied $k$ times resulting in $\ket{\psi}_k^X = \big(X_{90}^{\text{Hann}}\big)^k \ket{\psi}^X$. The Z projection value is measured after the k$^{\text{th}}$ X$_{90}^{\text{Hann}}$ gate yielding $ \langle \hat{Z}_k^X \rangle$.  The loss function is defined as the $L_1$ norm between the ideal Z projection values ($ \langle Z_1^X \rangle  = 1$, $ \langle Z_2^X \rangle  = 1/2$) and the measured value:  $L_x = \sum_k^M 1/M |\langle Z_k^X \rangle  - \langle \hat{Z}_k^X \rangle |$. In the following experiments, M is set to 2.\\
\indent We choose learning rate $a_t = a_0/t^{0.602}$ and perturbation size $c_t = c_0/t^{0.101}$  as proposed in the original SPSA paper\cite{spall1992multivariate}. These values have been used in a variety of recent experimental demonstrations\cite{kandala2017hardware, hou2019experimental}. The coefficients of the exponential moving average of gradient and squared gradient are set to $\beta_t = \beta_0/t^\lambda$ and a constant, respectively. 

\subsection{Visualizing optimization trajectories.} 
\begin{figure}[!htbbp]
\includegraphics[width=\linewidth, trim= 5cm 20cm 4cm 20cm,clip]{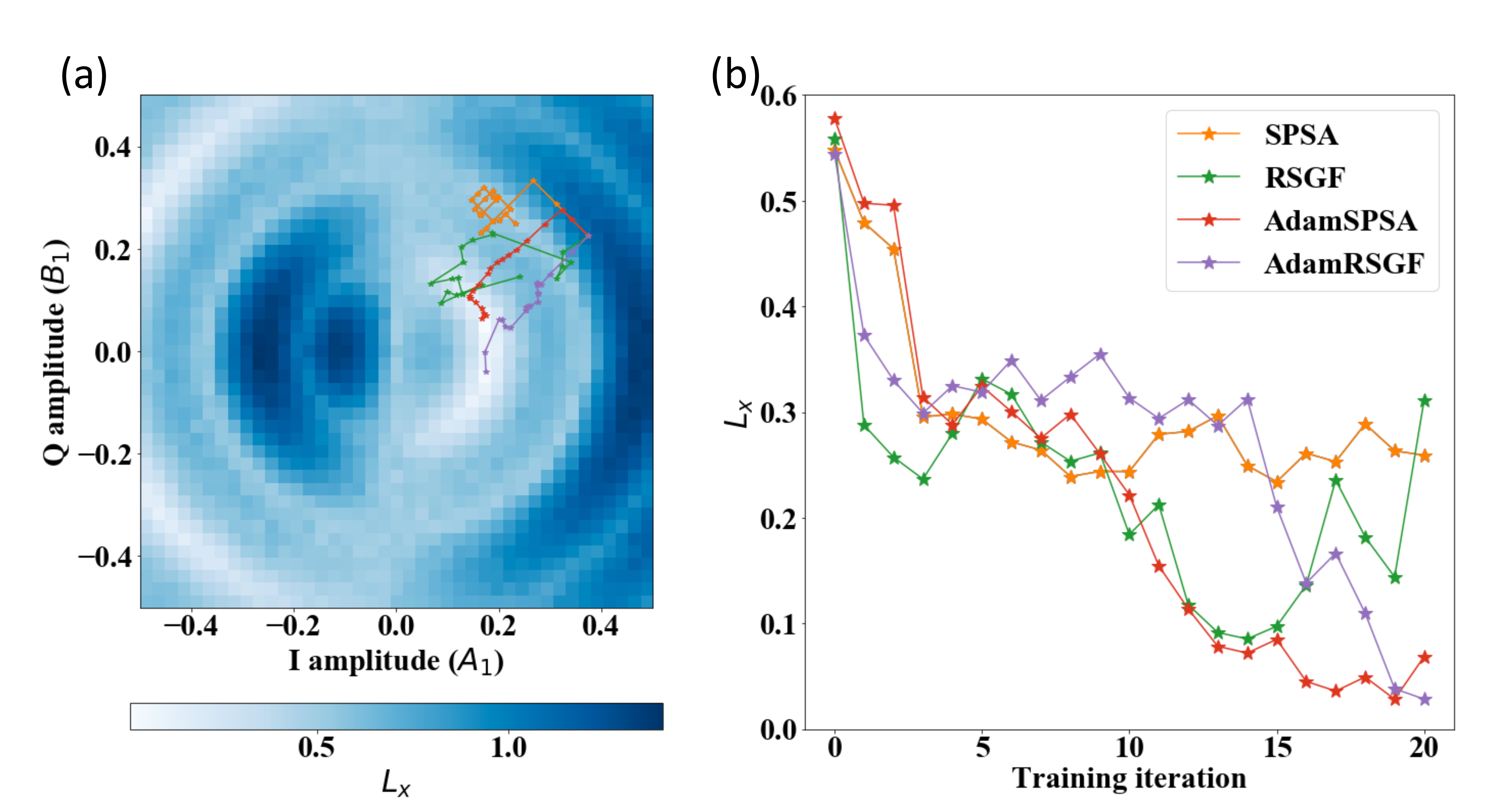}
\caption{\textbf{Optimization loss landscape}. (a) Optimization trajectories on a two-dimensional loss landscape. The loss function $L_x$ is measured by sweeping $A_1$ and $B_1$. For visualization purposes, $A_1$ and $B_1$ are arbitrarily initialized at (0.375, 0.225). Optimization trajectories for SPSA (blue), RSGF (yellow), AdamSPSA (green), and AdamRSGF (red) are measured independently. (b) Loss value at each training iteration. The SPSA algorithm (blue) gets trapped in a local minimum after six steps. The RSGF algorithm (yellow) makes a large update at step 15 and escapes out of the vicinity of the global minimum due to stochastic noise during function evaluation. Both AdamSPSA (green) and AdamRSGF (red) converge to the vicinity of the global minimum in twenty steps.}
\label{fig:fig1}
\end{figure}

For the purpose of visualizing the optimization trajectories, we optimize A$_1$ and B$_1$ and set the coefficients of higher frequency terms ($i > 1$) to 0. As shown in Figure \ref{fig:fig1}(a), the landscape is non-convex for two-variable optimization. When the initial condition is not in the vicinity of the global minimum, simple SPSA and RSGF methods get trapped in shallow local minima and suffer from fluctuations due to stochasticity of function evaluations. By contrast, AdamSPSA and AdamRSGF converge to the vicinity of the global minimum in twenty evaluations of function perturbations, as shown in Figure \ref{fig:fig1}(b).

\subsection{Benchmarking different algorithms.} 
\begin{figure}[!htbbp]
\includegraphics[width=\linewidth, trim= 7cm 32cm 8cm 16cm,clip]{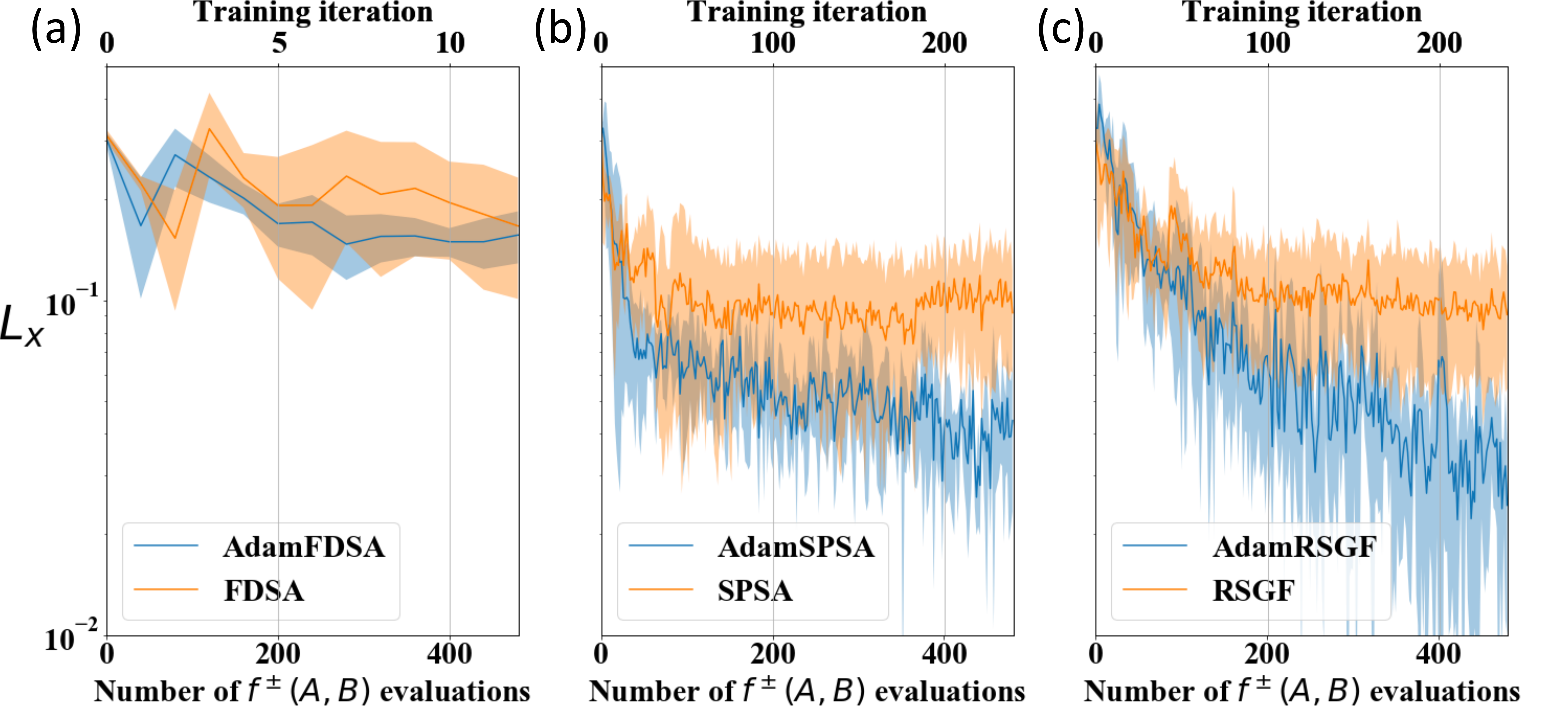}
\caption{\textbf{Benchmarking six algorithms on optimizing $\bm{L_x}$.} Loss function $L_x$ as a function of the number of $f^\pm(\bm{A}, \bm{B})$ evaluations for three different types of gradient estimations: FDSA (a), SPSA (b), and RSGF (c). Each gradient estimation method is tested with two different update rules: simple gradient descent and adaptive momentum estimation (Adam variant). Optimization starts from the same randomly generated initial point $(\bm{A}, \bm{B}) = (A_1, A_2, ... A_{10}, B_1, B_2, ..., B_{10})$. Each experiment contains 480 evaluations and is repeated five times. The mean of the five independent experiments is plotted as a solid line, and the standard deviation is shown as the shaded area in log scale.}
\label{fig:fig2}
\end{figure}

Here, we benchmark six algorithms: SPSA, RSGF, finite difference stochastic approximation (FDSA), AdamSPSA, AdamRSGF and the Adam variant of FDSA (AdamFDSA). The initial learning rate and perturbation size for all algorithms are the same: $a_0 = 0.032, c_0 = 0.016, \beta_0=0.999, \gamma_0=0.999, \lambda=0.4$. The first key observation is that the simultaneous perturbation based methods (SPSA, RSGF, AdamSPSA, AdamRSGF) converge faster than the finite difference methods (FDSA and AdamFDSA), as shown in Figure \ref{fig:fig2}. In high-dimensional domains, gradient estimation using finite difference is costly because computational cost scales linearly with dimension. For a domain of dimension twenty, each update requires 40 $f^\pm$ function evaluations. In 480 function evaluations, simple FDSA and AdamFDSA update parameters $(\bm{A}, \bm{B})$ only twelve times, whereas the other four random perturbation based algorithms can perform 240 gradient updates, greatly speeding up the convergence given the same learning rate coefficient $a_0$. \\
\indent We also see that, given an efficient gradient estimation method, using the Adam variant of the algorithm can further improve robustness. The loss trajectories for simple SPSA, RSGF, and FDSA have larger variances due to stochastic noise in the measurements and shallow local minima, as shown in Fig \ref{fig:fig2}. In addition, the losses in the Adam variants converge to a lower value (more than one standard deviation for both SPSA and RSGF). The performance improvement of AdamSPSA and AdamRSGF compared to simple SPSA and RSGF requires no additional function evaluations.

\subsection{Hann window high fidelity single qubit gate optimization.} 
\begin{figure}[!htbbp]
\includegraphics[width=\linewidth, trim= 0.5cm 0.2cm 1cm 0.2cm,clip]{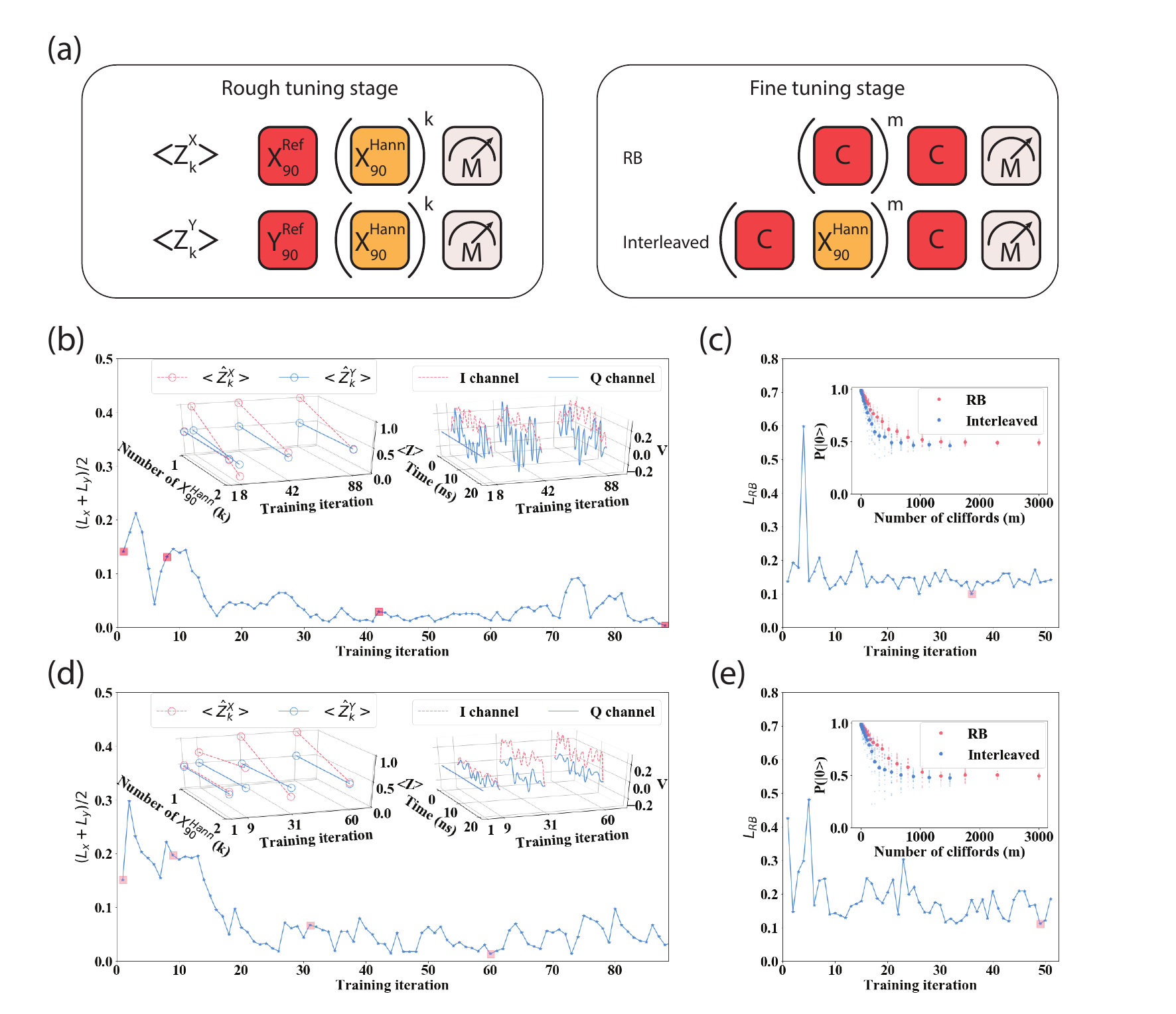}
\caption{\textbf{$\bm{X_{90}^{Hann}}$ gate optimization using the AdamSPSA and AdamRSGF algorithms.} (a) Gate optimization is composed of two stages: a rough tuning stage ((b), (d)) and a fine tuning stage ((c), (e)). In the rough tuning stage, the control pulse parameters are initialized to $(\bm{A}, \bm{B}) = \bm{0}_{1\times 20}$ and the loss function is defined as $L = 1/2(L_x + L_y) $. Inserts in (b) and (d) show the measured Z projection values $\langle \hat{Z}^X_k \rangle $, $\langle \hat{Z}^Y_k \rangle $  and corresponding waveforms in the I and Q channels at different training iterations (shown with red squares). In the fine tuning stage, the pulse parameters $(\bm{A}, \bm{B})$ are initialized at the lowest loss value from the first stage (training iteration 88 for AdamSPSA in (b), training iteration 60 for AdamRSGF in (d)). The loss function for the fine tuning stage is $L_{RB}$, which is extracted from randomized benchmarking (RB) experiments. Inserts in (c) and (e) show the RB and interleaved RB trajectories evaluated at Clifford numbers m = [0, 1, 2, 3, 4, 5, 6, 8, 11, 14, 19, 25, 32, 42, 55, 72, 93, 122, 159, 208, 272, 355, 463, 605, 790, 1032, 1347, 1759, 2297, 3000]. Interleaved RB is measured using the pulse parameters $(\bm{A}, \bm{B})$ at the lowest loss value found in the fine tuning stage (shown with red squares) to extract gate fidelities of $0.9993 \pm 5$E$-5$ for AdamSPSA and $0.9993 \pm 10$E$-5$ for AdamRSGF.}
\label{fig:fig3}
\end{figure}

To verify that AdamSPSA and AdamRSGF are suitable for optimizing a quantum system with high precision, we use both methods to perform single-qubit $X_{90}^{\text{Hann}}$ gate tune ups and achieve fidelities over 99.9\%.  The gate optimization contains two stages: A rough tuning stage and a fine tuning stage, shown in Fig. \ref{fig:fig3}(a). \\
\indent In the rough tuning stage, to enforce the rotation of $X_{90}^{\text{Hann}}$ along the $x$-axis, we introduce a reference Y loss $L_y$ in addition to the benchmark setup from the previous section. To obtain $L_y$, the qubit is initialized in $\ket{\psi}^Y = 1/\sqrt{2} \ket{0} + 1/\sqrt{2} \ket{1}$ by applying a reference Y$_{90}^{\text{Ref}}$ gate. Additional X$_{90}^{\text{Hann}}$ gates are applied k times resulting in $\ket{\psi}^Y_k = \big(X_{90}^{\text{Hann}}\big)^k \ket{\psi}^Y$. The Z projection value is measured after the k$^{th}$ X$_{90}^{\text{Hann}}$ gate, yielding $ \langle \hat{Z}^Y_k \rangle $.  The Y loss function is defined as $L_y = \sum_k^M 1/M |\langle Z^Y_k \rangle  - \langle \hat{Z}^Y_k \rangle |$, which is the $L_1$ norm of the difference between the ideal Z projection value $ \langle Z^Y_k \rangle  = 0.5$ and the measured value $ \langle \hat{Z}_k \rangle $. The total loss used in optimization is the average of the reference X and Y loss: $L = 1/2(L_x+L_y)$. As shown in Fig. \ref{fig:fig3}(b) and (d), AdamSPSA and AdamRSGF are able to find waveforms for the X$_{90}^{\text{Hann}}$ gate with loss to close to zero.\\
\indent In the fine tuning stage, we employ Clifford-based randomized benchmarking (RB) to further improve the gate fidelity \cite{kelly2014optimal, rol2017restless}. We choose the initial condition for $(\bm{A} ,\bm{B})$ in this stage to be the parameters with the lowest loss value in the previous stage.  The Clifford gates are constructed from $\{I, X_{90}^{\text{Hann}}, X_{-90}^{\text{DRAG}}, X_{180}^{\text{DRAG}}$ $, Y_{\pm90}^{\text{DRAG}}\}$, where 20\,ns DRAG pulses with 6\,ns buffer between two pulsese are tuned up using the standard calibration\cite{sheldon2016characterizing}. Here, we set $\lambda=0.1$. For AdamSPSA, we set $\alpha_0 = 0.002, c_0 = 0.002$, and for AdamRSGF, we choose $\alpha_0 = 0.004$ and $c_0 = 0.004$. We measure the Z-projection $|\langle \hat{Z} \rangle |$ as a function of the number of applied Clifford gates m, and then we extract the average Clifford fidelity $p$ by fitting to $|\langle \hat{Z} \rangle | = Ap^m + B$. The loss function is defined as the average Clifford infidelity $L_{RB} = (1-p) \times 100$. As shown in Fig \ref{fig:fig3} (c) and (e), both AdamRSGF and AdamSPSA further improve the gate fidelity via noisy gradient estimation from $L_{RB}$. To evaluate the $X_{90}^{\text{Hann}}$ gate fidelity, interleaved RB is performed for each set of optimized $(\bm{A} ,\bm{B})$ values. Here, $X_{90}^{\text{Hann}}$ is appended to each Clifford gate \cite{magesan2012efficient}, as shown in Fig\ref{fig:fig3} (a).  The resulting $X_{90}^{\text{Hann}}$ gate fidelities are both 0.9993 ($\pm 5e^{-5}$ for AdamSPSA and $\pm 10e^{-5}$ for AdamRSGF), while the coherence-limited fidelity is 0.9993.
\section*{Discussion}
\indent \indent In this paper, we have shown that incorporating momentum and adaptive learning rate scheduling into simultaneous-perturbation-based algorithms significantly improves quantum optimization results for high-dimensional non-convex problems compared to the finite difference based algorithm and two random-perturbation-based algorithms. We benchmarked our new algorithms, AdamSPSA and AdamRSGF, using quantum optimal control as a subset of quantum optimization. Without using any prior knowledge of signal distortion from classical electronics or weak anharmonicity of the qubit, our algorithms enabled efficient learning of Hann window pulses that achieve state-of-the-art gate fidelity. While previous studies have focused on reducing the time per function evaluation, we have designed and demonstrated optimization algorithms that estimate gradient efficiently and update parameters robustly. These results open up exciting new opportunities in bridging optimization and quantum optimal, such as fully automatic feedback-based gate optimization in multi-qubit processors from trivial initial conditions and in optimal control for different types of qubits such as Fluxonium and 0-$\pi$ \cite{manucharyan2009fluxonium, groszkowski2018coherence, di2019control}. Furthermore, our algorithms can be broadly applied to other quantum optimization problems such as quantum chemistry and quantum machine learning in NISQ systems as well as black-box optimization problems outside the quantum world \cite{cao2018quantum,biamonte2017quantum}.
\section*{Methods}
\indent \indent The transmon qubit was made using bridge-free double-angle evaporation \cite{potts2001cmos, koch2007charge}. The coplanar waveguide cavity for qubit readout and control was fabricated on a 200-nm film of niobium sputtered on a 500$\mu m$ thick sapphire substrate using standard optical lithography and dry etching techniques.\\
\indent Qubit control waveforms were generated on a classical computer and sent to the qubit via input lines (details in the supplement). The initial learning rate and step size $a_0, c_0$ were optimized for the basic versions of the SPSA and RSGF algorithms by performing a random search on logarithmic scale with A = [0.064, 0.032, 0.016, 0.008, 0.004], C = [0.032, 0.016, 0.008, 0.004] before benchmarking. All other algorithms directly adopted the same $a_0, c_0$ without further tuning. The momentum decay rate $\lambda$ was optimized by linear scale grid search from 0 with step size 0.1. In the fine tuning stage, initial leaning rate and step size $a_0, c_0$ were optimized by logarithmic search on (A, C) = [(0.032, 0.032), (0.016, 0.016), (0.008, 0.008), (0.004,0.004), (0.002, 0.002)] while keeping $\lambda$ constant.
\section*{References}
\bibliographystyle{naturemag}
\bibliography{citation}

\section*{Acknowledgements}
\indent \indent This work was supported by IARPA under contract W911NF-16-1-0114-FE. The authors would like to acknowledge Christie Chiu, Andr\a'as Gyenis, Anjali Premkumar, Basil Smitham and Sara Sussman for valuable comments on the manuscript. Device was fabricated in the Princeton University Quantum Device Nanofabrication Laboratory and the Princeton Institute for the Science and Technology of Materials.
\section*{Author Contributions}
\indent \indent Z.L. designed and fabricated the sample, proposed and proved the algorithms, and performed the data analysis. Z.L., P.M. performed the measurements. S.G. supervised the optimization algorithms. A.H. supervised the whole process. All authors contributed to the preparation of this manuscript.
\section*{Competing financial interests}
\indent \indent The authors declare no competing financial interests.

\newpage

\newpage

\newpage

\newpage

\end{document}


\title{Supplementary materials}

\author{Zhaoqi Leng$^1$, Pranav Mundada$^2$,  Saeed Ghadimi $^3$ \& Andrew Houck $^2$}

\maketitle
\begin{affiliations}
 \item Department of Physics, Princeton University.
 \item Department of Electrical Engineering, Princeton University.
 \item Department of Operations Research and Financial Engineering, Princeton University.
\end{affiliations}

\section{Device parameters}
The Hamiltonian for a transmon qubit is 
\begin{equation}
\begin{split}
H/\hbar = \omega a^\dagger a - \frac{\alpha}{2} a^\dagger a^\dagger a a.
\end{split}
\end{equation}
The transmon has frequency $\omega/2\pi= 5.16$ GHz and anharmonicity $\alpha = 320$ MHz. The relaxation time is $T_1 = 32 \mu s$ . The coherence time measured by Ramsey and Echo experiments are $T_{2R} = 26 \mu s$ and $T_{2E} = 40 \mu s$ respectively. 
\section{Pulse shape distortion}
\begin{figure}[h!]
\includegraphics[width=\textwidth, trim= 0cm 6cm 0cm 4cm,clip]{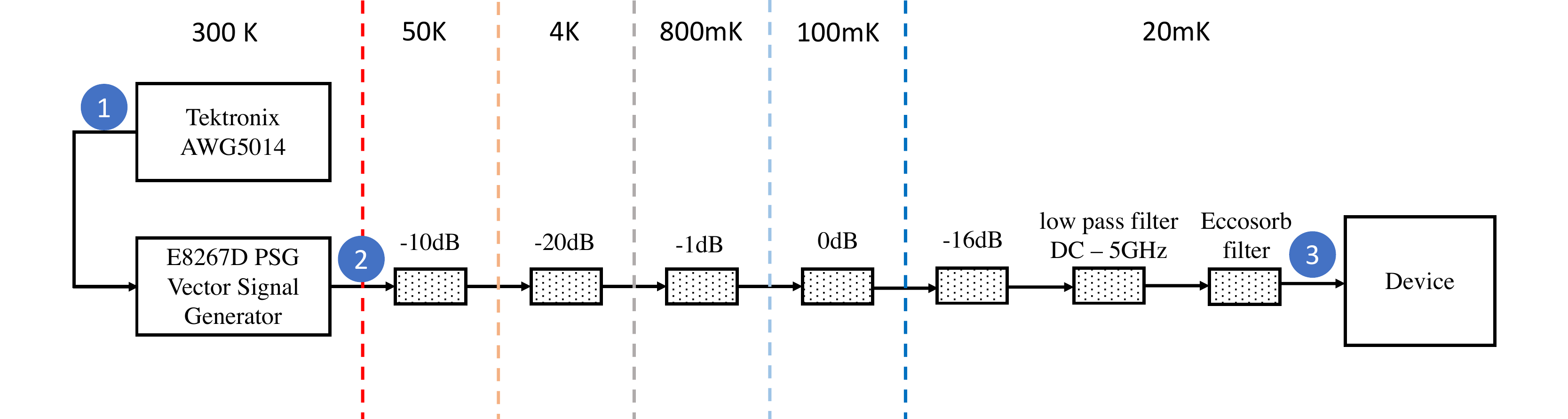}
\caption{\textbf{Schematic of input setup}. The control signal is generated by room-temperate electronics and goes through the input line until it reaches the device at around 20mK. To illustrate the signal distortion, waveforms are measured at three different locations (1), (2) and (3) at room temperature.}
\label{fig:fig1}
\end{figure}

\begin{figure}[h!]
\includegraphics[width=\textwidth, trim= 18cm 27cm 19cm 30cm,clip]{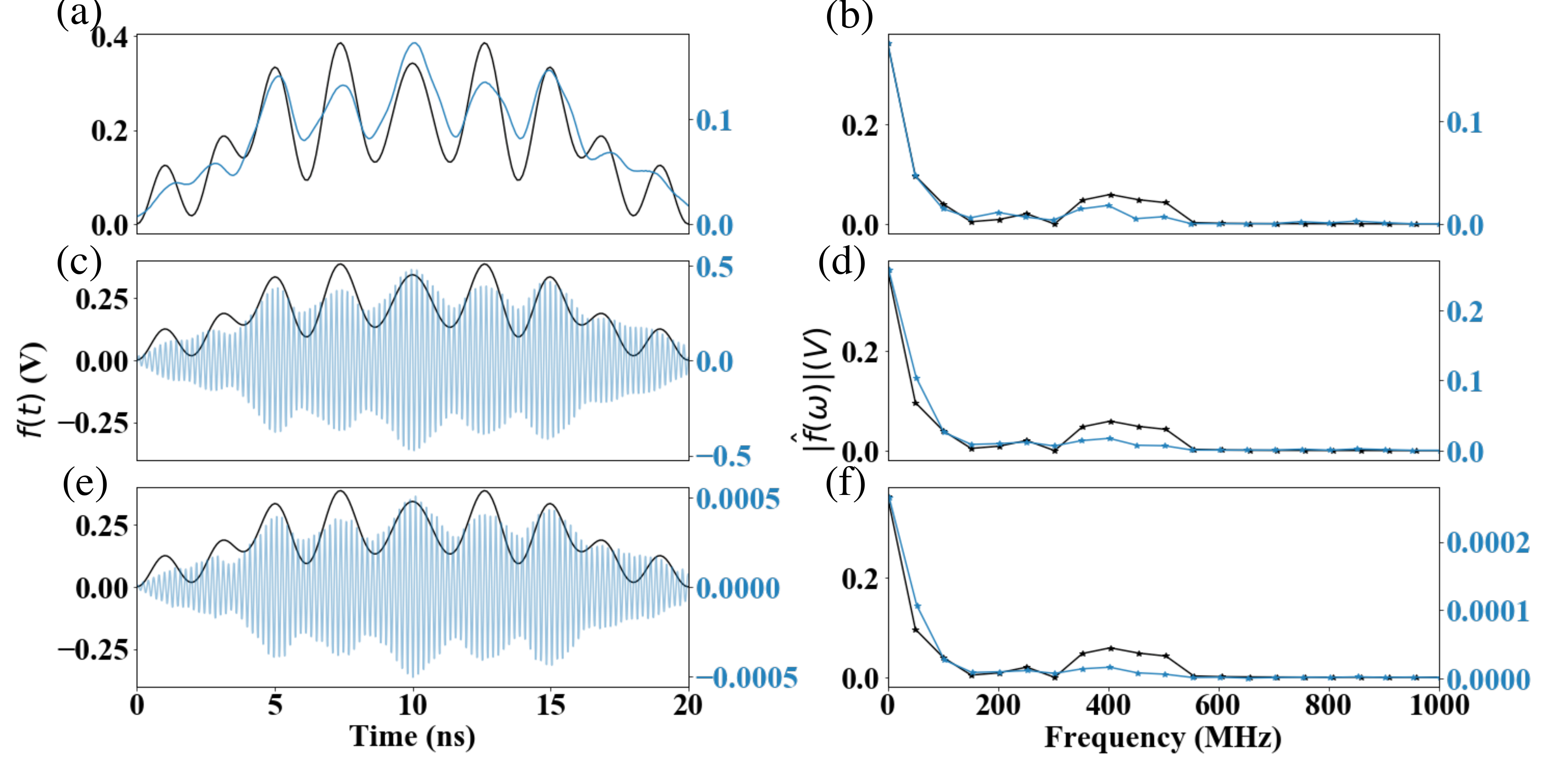}
\caption{\textbf{Qubit waveform at different stages}. In (a), (c), (e), the black lines represent the optimal waveform $f(t)$ generated on a computer. The blue lines in (a), (c), (e) are measured waveforms at location (1), (2), (3) in Fig 1. The blue and black lines in (b), (d), (f) represent the amplitudes of Fourier components $|\hat{f}(\omega)|$ in plot (a), (c), (e), where the frequency of the blue lines in (d) and (f) are shifted by $\omega_0/2\pi = 5.16 GHz$.}
\label{fig:fig2}
\end{figure}

The qubit control waveform is defined on a classical computer and sent into the device via input lines, shown in Fig \ref{fig:fig1}.
Before reaching the device that hosts the quantum chip, the control waveform generated on a classical computer is distorted by classical electronics such as waveform generators, signal generators, line attenuations etc. Here, we use the optimal waveform for the $X_{90}^{Hann}$ gate, shown in the main text Fig 4 (c), as a demonstration. The window function of the optimal waveform is sent directly from a computer to an arbitrary waveform generator. As shown in Fig \ref{fig:fig2} (a),(b) , the waveform is distorted after the first classical electronics: the waveform gets attenuated less near 200MHz and more around 400MHz. Further down the line, the window function is distorted after mixing with high frequency signals from a vector signal generator and by other classical electronics in the fridge, as shown in Fig \ref{fig:fig2} (c), (d), (e), and (f).

\section{Convergence for AdamSPSA and AdamRSGF}
For convex optimization $f(\bm{\theta}^*) := \text{inf} f(\bm{\theta})$, Kushner and Clark's (KC) condition is a necessary and sufficient condition for $\lim_{t\to \infty}\bm{\theta}_t = \bm{\theta}^*$, where the relevant conditions for proving asymptotic convergence for the Adam version of SPSA and RSGF in convex settings are listed below\cite{wang1996equivalent, spall1992multivariate}:
\begin{enumerate}[start=1,label={(\bfseries A\arabic*):}]
    \item \{$a_t$\} is a sequence of positive real number and $\lim_{t\to\infty} a_t = 0$ and $\sum_{t=1}^{\infty}a_t = \infty$ \label{conv1},
    \item $\lim_{t\to \infty} \bm{b}_t = 0$ \label{conv2},
    \item $\lim_{n\to \infty} ( \underset{n \leq p \leq m(n, T)}{\text{sup}}||\sum_{t=n}^{p} a_t \bm{e}_t||) = 0$ (KC condition), \label{conv3}
\end{enumerate}
where $a_t$ is the learning rate, $\bm{b}_t$ is the biased of the gradient, and $\bm{e}_t$ is the stochastic noise of the gradient as defined in the main text Eqn (1). Note that both $\bm{b}_t$ and $\bm{e}_t$ are $p$ dimensional vectors as the parameter $\bm{\theta}_t$. Here, we assume the function gradient, measurement induced noise, and the function evaluation are bounded by $||\bm{g}_t^i|| \leq \alpha_0$, $||\epsilon^\pm||\leq \alpha_1$ and $||f^\pm|| \leq \alpha_2$. The inverse random perturbation for (Adam)SPSA is bounded by  $||1/\bm{\Delta}_t||\leq \alpha_3$, and for (Adam)RSGF, the expectation value of random perturbation is bounded by E$||\bm{u}_t||\leq \alpha_3$. The bias is bounded by $||\bm{b}_t||\leq \alpha_4 c_t^2$ as the bias is O($c_t^2$) \cite{spall1992multivariate, ghadimi2013stochastic}.\\
The KC condition can be satisfied when $\sum_{t=1}^{\infty} (a_t/c_t)^2 < \infty$ as shown in the original paper on SPSA \cite{spall1992multivariate}, where for $a_t = a_0/t^\alpha$ and $c_t = c_0/t^\zeta$, $\alpha - \zeta >0.5$.
\subsection{Momentum SPSA/RSGF}
When $\sqrt{\hat{\bm{v}}_t}$ is set to be a constant, the Adam SPSA/RSGF become momentum SPSA/RSGF where the learning rate is no longer adaptive based on the history of squared gradients.
Since the bias correction coefficients for the moving averages of gradients and squared gradients converge to 1 as $\lim t\to \infty$, the following proof do not consider the bias correction for simplicity. The update rule for momentum update without adjusting the bias from exponential moving average can be written as the following:
\begin{equation}
\begin{split}
\hat{\bm{\theta}}_{t+1} &= \hat{\bm{\theta}}_{t}  - a_t \bm{m}_t,\\
\bm{m}_t &= \beta_t \bm{m}_{t-1} + (1 - \beta_t)\hat{\bm{g}}_t.
\end{split}
\end{equation}
Different from the gradient descent $\hat{\bm{\theta}}_{t+1} = \hat{\bm{\theta}}_{t}  - a_t \hat{\bm{g}}_t$, the momentum term $\beta_t \bm{m}_{t-1}$ introduces additional bias and noise,
\begin{equation}
\begin{split}
\beta_t \bm{m}_{t-1} &= \beta_t(\beta_{t-1} \bm{m}_{t-2} + (1 - \beta_{t-1})\hat{\bm{g}}_{t-1})\\
& = \Pi_{i=0}^{t-1}\beta_i \bm{m}_0 + \sum_{N=0}^{t-1} (1 - \beta_{t-1-N})\hat{\bm{g}}_{t-1-N}\Pi_{i=0}^N \beta_{t-i}\\
& = \sum_{N=0}^{t-1} (1 - \beta_{t-1-N})(\bm{g}_{t-1-N} +  \bm{b}_{t-1-N} + \bm{e}_{t-1-N})\Pi_{i=0}^N \beta_{t-i}.
\end{split}
\end{equation}
As a result, $\beta_t \bm{m}_{t-1}$ can be split into two terms, a bias term and a zero mean noise term. 
\begin{equation}
\begin{split}
\text{Bias: } \sum_{N=0}^{t-1}& (1 - \beta_{t-1-N})(\bm{g}_{t-1-N} +  \bm{b}_{t-1-N})\Pi_{i=0}^N \beta_{t-i}, \\
\text{Noise: } \sum_{N=0}^{t-1}& (1 - \beta_{t-1-N}) \bm{e}_{t-1-N}(\theta)\Pi_{i=0}^N \beta_{t-i}.
\end{split}
\end{equation}
The additional bias term from Eqn (4) is bounded by:\\
\begin{equation}
\begin{split}
\label{eq: bias}
\lim_{t\to \infty} &\sum_{N=0}^{t-1} ||(1 - \beta_{t-1-N})(\bm{g}_{t-1-N} +  \bm{b}_{t-1-N})\Pi_{i=0}^N \beta_{t-i} || \\
\leq  \lim_{t\to \infty} & \sum_{N=0}^{t-1} (||\bm{g}_{t-1-N}|| +  ||\bm{b}_{t-1-N}||)\Pi_{i=0}^N \beta_{t-i}\\
\leq \lim_{t\to \infty} & (\alpha_0p + \alpha_4 c^2_{t-1-N})\sum_{N=0}^{t-1} \Pi_{i=0}^N \beta_{t-i}.
\end{split}
\end{equation}
The contribution of the additional noise term from Eqn (4) to $\lim_{n\to \infty} ( \underset{n \leq p \leq m(n, T)}{\text{sup}}||\sum_{t=n}^{p} a_t \bm{e}_t||)$ can be rewrite as the following by Doob's martingale inequality:
\begin{equation}
\begin{split}
\label{eq: nosie}
\lim_{t\to \infty}E||\sum_{t=n}^p & a_t \sum_{N=0}^{t-1} (1 - \beta_{t-1-N})\bm{e}_{t-1-N}\Pi_{i=0}^N \beta_{t-i} ||\\
\leq \lim_{t\to \infty}  \sum_t^p &a_t  \sum_{N=0}^{t-1} (1 - \beta_{t-1-N}) \Pi_{i=0}^N (\beta_{t-i} ) E||\bm{e}_{t-1-N}||\\
\leq \lim_{t\to \infty} \sum_t^p &a_t  \sum_{N=0}^t \Pi_{i=0}^N (\beta_{t-i} ) 2p (\alpha_1 + \alpha_0) \alpha_3 c_{t-1}^{-1}\\
\leq \lim_{t\to \infty} \sum_t^p &a_t  2p (\alpha_1 + \alpha_2) \alpha_3 c_{t-1}^{-1} \sum_{N=0}^{t-1} \Pi_{i=0}^N (\beta_{t-i} ).\\
\end{split}
\end{equation}
If $\beta_t$ is defined as an annealing function $\beta_t = 1/t^{\lambda}$ similar to $a_t$ and $c_t$, then we define $S_t$
\begin{equation}
\begin{split}
S_t &= \lim_{t\to \infty} \sum_{N=0}^{t-1} \Pi_{i=0}^N \beta_{t-i} = \lim_{t\to \infty} \sum_{N=0}^{t-1} \Pi_{i=0}^N \frac{1}{(t-i)^\lambda}\\
S_{t+1} &= \lim_{t\to \infty} \sum_{N=0}^{t} \Pi_{i=0}^N \frac{1}{(t+1-i)^\lambda}
\end{split}
\end{equation}
Since $S_{t+1} (1+t)^\lambda = 1 + S_t$,
\begin{equation}
\begin{split}
\lim_{t\to \infty} \sum_{N=0}^{t-1} \Pi_{i=0}^N \beta_{t-i} = \lim_{t\to \infty} \frac{1}{t^\lambda - 1}.
\end{split}
\end{equation}
Therefore, $\lambda > 0$ and $\lambda + \alpha - \zeta > 1$ are necessary conditions to ensure (A2) and (A3).\\
On the other hand, a trivial condition to satisfy both (A2) and (A3) is $\beta_t = 0$ for $t>M$, where $M$ is a finite integer. This is equivalent to convert Momentum SPSA and RSGF to the original SPSA and RSGF after step $M$, which relaxes the condition for $\lambda$ from $\lambda + \alpha - \zeta > 1$ and $\lambda > 0$ to any $\lambda$ value for $t\leq M$, as long as $\beta_t = 0$ for $t>M$. \\
\subsection{Adam SPSA/RSGF}
We can absorb the adaptive term for the learning rate by defining an effective learning rate $\hat{\bm{a}}_t^i = a_t/(\sqrt{\hat{\bm{v}}_t^i} + \delta)$, where $\hat{\bm{a}}_t$ is a $p$ dimensional vector. Additional constrains are needed for $\hat{\bm{a}}_t^i$ to satisfy (A1) and (A3).\\
The adaptive learning rate is bounded by $\hat{\bm{a}}_t^i \leq  a_t/\delta$, thus $\lim_{t\to\infty} \hat{\bm{a}}_t^i \leq \lim_{t\to\infty} a_t/\delta = 0$.\\
The moving average of the squared gradient is \\
\begin{equation}
\begin{split}
\bm{v}_{t}^i &= \gamma_{t} \bm{v}_{t-1}^i + (1 - \gamma_{t})(\hat{\bm{g}}^i)^2_{t-1}\\
& = \sum_{N=1}^{t-1} (1 - \gamma_{t-1-N}) (\hat{\bm{g}}_{t-1-N}^i)^2\Pi_{i=0}^N \gamma_{t-i} + (1 - \gamma_{t})(\hat{\bm{g}}^i)^2_{t-1}.\\
\end{split}
\end{equation}
Here, we assume $\gamma_t \in (0, 1)$ is a constant, thus
\begin{equation}
\begin{split}
\lim_{t \to \infty}\bm{v}_{t}^i &=  \lim_{t \to \infty}(1 - \gamma)\sum_{N=1}^{t-1} (\hat{\bm{g}}^i_{t-1-N})^2 \gamma^{N+1} + \lim_{t \to \infty}(1 - \gamma)(\hat{\bm{g}}^i_{t-1})^2\\
& \leq  \lim_{t \to \infty}(1 - \gamma) (2(\alpha_1 + \alpha_2) \alpha_3 c_{t-1})^2\sum_{N=1}^{t-1}  \gamma^{N+1} + \lim_{t \to \infty}(1 - \gamma)(2(\alpha_1 + \alpha_2)\alpha_3 c_{t-1})^2\\
&=(2(\alpha_1 + \alpha_2)\alpha_3 c_{t-1}^{-1})^2.
\end{split}
\end{equation}
For AdamRSGF, $||\bm{\mu}_t||$ is not bounded by $\alpha_3$. Therefore, additional cutoff need to be imposed to ensure  $||\bm{\mu}_t|| < \alpha_3$. The additional cutoff is not a general propriety of RSGF method. If we use smoothing function such as uniform distribution over the surface of $l_2$  ball of radius $\sqrt{p}$, where $||\bm{\mu}_t||$ is bounded, no additional cutoff is necessary \cite{duchi2015optimal}. \\
To ensure $\sum_{t=1}^{\infty}\hat{\bm{a}}_t^i = \infty$ with $a_t = a_0/t^\alpha$, $c_t = c_0/t^\zeta$:
\begin{equation}
\begin{split}
\sum_{t=1}^{\infty} a_t/(\sqrt{\bm{v}_t^i} + \delta) \geq &\sum_{t=1}^{\infty} a_t/(2(\alpha_1 + \alpha_2)\alpha_3c_{t-1}^{-1} + \delta )  \\
 \geq &\sum_{t=1}^{\infty} a_t c_{t-1}/(2(\alpha_1 + \alpha_2)\alpha_3+1)\\
 > & \sum_{t=1}^{\infty} \frac{1}{t^{\alpha + \zeta}}a_0c_0/(2(\alpha_1 + \alpha_2)\alpha_3+1).
\end{split}
\end{equation}
The p-series diverges when $\alpha + \zeta \leq 1$ \\
(A3) is satisfied due to
\begin{equation}
\begin{split}
\lim_{n\to \infty} &( \underset{n \leq p \leq m(n, T)}{\text{sup}}||\sum_{t=n}^{p} \hat{\bm{a}}_t^i \bm{e}_t^i||) \\
\leq \lim_{n\to \infty}& ( \underset{n \leq p \leq m(n, T)}{\text{sup}}||\sum_{t=n}^{p} a_t  \bm{e}_t^i||)/\sqrt{\delta}.\\
\end{split}
\end{equation}

\bibliographystyle{naturemag}
\bibliography{supplement}